\documentclass[runningheads]{llncs}
\usepackage[T1]{fontenc}
\usepackage{graphicx}
\usepackage[scientific-notation=false]{siunitx}
\usepackage[acronym]{glossaries}
\newacronym{pe}{PE}{processing element}
\newacronym{gpu}{GPU}{graphic processing unit}
\newacronym{sm}{SM}{streaming multiprocessor}
\newacronym{cuda}{CUDA}{compute unified device architecture}
\newacronym{mpi}{MPI}{message passing interface}
\newacronym{openmp}{OpenMP}{open multi-processing}
\newacronym{ai}{AI}{artificial intelligence}
\newacronym{risc-v}{RISC-V}{reduced instruction set computer V}
\newacronym{tcdm}{TCDM}{tightly coupled data memory}
\newacronym{noc}{NoC}{nework on chip}
\newacronym{pram}{PRAM}{parallel random access machine}
\newacronym{numa}{NUMA}{non uniform memory access}
\newacronym{csr}{CSR}{control status register}

\newacronym{gemm}{MATMUL}{\textit{matrix-matrix multiplication}}
\newacronym{dotp}{DOTP}{\textit{dot-product}}
\newacronym{axpy}{AXPY}{\textit{ax plus y}}
\newacronym{dct}{DCT}{\textit{direct cosine transform}}
\newacronym{conv2D}{Conv2D}{\textit{2D-convolution}}
\newacronym{fft}{FFT}{\textit{fast Fourier transform}}

\newacronym{ofdm}{OFDM}{Orthogonal Frequency Division Multiplexing}
\newacronym{push}{PUSCH}{Physical Uplink Shared Channel}
\newacronym{phy}{PHY}{Physical Layer}
\newacronym{mimo}{MIMO}{Multiple-Input Multiple-Output}
\newacronym{3gpp}{3GPP}{3rd Generation Partnership Project}

\newacronym{cdf}{CDF}{cumulative distribution function}
\newacronym{api}{API}{application programmable interface}
\newacronym{rtl}{RTL}{register transfer level}
\newacronym{wfi}{WFI}{wait for interrupt}
\newacronym{ipc}{IPC}{instructions per cycle}
\newacronym{sfr}{SFR}{synchronization free region}
\usepackage{bibnames}
 \usepackage{cite}
\begin{document}
\title{Fast Shared-Memory Barrier Synchronization for a 1024-Cores RISC-V Many-Core  Cluster}%\thanks{Supported by Huawei Sweden AG.}}
\titlerunning{Shared-Memory Barrier Synchronization for a 1024-Cores RISC-V Cluster}
%\author{Authors' names were omitted for double-blind review}
%\institute{}
\author{Marco Bertuletti\inst{1}\orcidID{0000-0001-7576-0803} \and
Samuel Riedel\inst{1}\orcidID{0000-0002-5772-6377} \and
Yichao Zhang\inst{1}\orcidID{0009-0008-7508-599X} \and Alessandro Vanelli-Coralli\inst{1,2}\orcidID{0000-0002-4475-5718} \and Luca Benini\inst{1,2}\orcidID{0000-0001-8068-3806}}
\authorrunning{M. Bertuletti et al.}
\institute{ETH Z\"{u}rich, R\"{a}mistrasse 101, 8092 Z\"{u}rich, Switzerland \email{\{mbertuletti,sriedel,yiczhang,avanelli,lbenini\}@iis.ee.ethz.ch} \and Universit\`a di Bologna, via Zamboni 33, 40126 Bologna, Italy\\
}
\maketitle              % typeset the header of the contribution
\begin{abstract}
Synchronization is likely the most critical performance killer in shared-memory parallel programs. With the rise of multi-core and many-core processors, the relative impact on performance and energy overhead of synchronization is bound to grow. This paper focuses on barrier synchronization for \textit{TeraPool}, a cluster of 1024 RISC-V processors with non-uniform memory access to a tightly coupled 4MB shared L1 data memory. We compare the synchronization strategies available in other multi-core and many-core clusters to identify the optimal native barrier kernel for \textit{TeraPool}. We benchmark a set of optimized barrier implementations and evaluate their performance in the framework of the widespread fork-join Open-MP style programming model. We test parallel kernels from the signal-processing and telecommunications domain, achieving less than 10\% synchronization overhead over the total runtime for problems that fit \textit{TeraPool}'s L1 memory. By fine-tuning our tree barriers, we achieve $1.6\times$ speed-up with respect to a naive central counter barrier and just 6.2\% overhead on a typical 5G application, including a challenging multistage synchronization kernel. To our knowledge, this is the first work where shared-memory barriers are used for the synchronization of a thousand processing elements tightly coupled to shared data memory.

\end{abstract}
\section{Introduction}
With the slow-down of Moore's Law at the turning of the century, multi-core systems became widespread, to sustain performance increase at an acceptable power budget in a scenario of diminishing returns for technology scaling~\cite{Theis_moorelaw_2017}. Nowadays, the popularity of many-core systems increases, as they offer huge parallel computing power to achieve top performance on embarrassingly parallel workloads, from genomics over computational photography and machine learning to telecommunications~\cite{Li_Bioninform_2010, Muralidhar_ACM_2022, Venkataramani_ACM_2020}. For example, NIVIDIA's \textit{H100 Tensor Core} \gls{gpu} features \num{144} \glspl{sm} with \num{128} \glspl{pe} \cite{Hopperarchitecture_NVIDIA_2022}, \gls{ai} accelerators such as \textit{TsunAImi} assemble \num{4} \textit{RunAI200} chiplets \cite{Runai200_2022} with \num{250000} \glspl{pe} within standard SRAM arrays for an at-memory design, \textit{Esperanto}'s \textit{ET-Soc-1} chip has over a thousand RISC-V processors on a single chip and more than \num{160} MB of on-chip SRAM\cite{Ditzel_Esperanto_HotChips_2021}. 

As the number of cores increases, scaling up architectures by instantiating many loosely-coupled clusters is a common architectural pattern adopted by many-cores to ensure modularity. However, this approach introduces overheads, including workload distribution, data allocation and splitting, inter-cluster communication and synchronization. To reduce these costs, increasing the cluster size is therefore desirable, as a direction space exploration the physical viability of this direction was demonstrated by \textit{MemPool}~\cite{Mempool_DATE_2020,Mempool_Journal_2023}, which couples \num{256} RISC-V Snitch\cite{Snitch_IEEE_2021} \glspl{pe} to a shared data memory ensuring low access latency. In this paper we further scale-up \textit{MemPool} and increase the core count to \num{1024}.

\textit{MemPool} and \textit{TeraPool} have a fork-join (the abstraction at the base of Open-MP) programming model: sequential execution forks to a parallel section, where \glspl{pe} access concurrently the shared memory. Barriers are used to synchronize and switch back to the sequential execution. The cost of barrier synchronization is a relevant factor to determine the performance of a parallel program and scales with the number of \glspl{pe} involved~\cite{Mellor_1991_ACM_TCS}. Despite the high core count, the synchronization overhead in \textit{TeraPool} must be minimal, as we desire speedup also for kernels that do not have lots of work in each parallel thread. Moreover, synchronizing only some cores in the cluster must be possible, to ease workload distribution, increase the \glspl{pe} utilization, and ensure more degrees of freedom in the parallel decomposition of a full application.

In this paper, we challenge the fork-join programming model, implementing fast barriers for a shared-memory tightly coupled cluster of 1024 cores. Our contributions include:
\begin{itemize}
    \item A comparison of the synchronization strategies adopted on other large-scale many-core systems, with a focus on how the hierarchy of the hardware architecture affects the barrier implementation.
    \item The implementation of a central counter barrier and a k-ary tree synchronization barrier for \textit{TeraPool}, exploiting hardware support to trigger the wakeup of all the \glspl{pe} in the cluster or a fraction of them.
    \item An in-depth analysis of the performance of \textit{TeraPool} on shared-memory parallel kernels synchronized with the implemented barriers, showing that the granularity of synchronization can be tuned on the basis of the kernel characteristics and that the barrier selection is an important stage of the kernel optimization.
\end{itemize}

A key insight is that tree barriers give up to $1.6\times$ speed-up with respect to centralized barriers and less than 6.2\% impact of synchronization on the total runtime of a full application from the field of 5G communications. Focusing on a RISC-V-based open-source many-core system enables us to provide an in-depth analysis of the synergies between our barrier implementations and the underlying hardware architecture. This is a key advantage over proprietary vendor-specific solutions, where an incomplete disclosure of architecture prevents effective hardware-software co-design.

\section{Related work}
In the following, we survey relevant contributions on the synchronization of tens to hundreds of \glspl{pe}. Most of the implementations focused on cached-based systems, with \glspl{pe} grouped in hierarchies and sharing only the last-level cache. 

In \cite{Mohamed_2023_Wiley}, an extended butterfly barrier was designed for Intel Xeon Phi \textit{Knight's Landing} processor, whose \num{72} \glspl{pe}, each having a private L1 data cache, are grouped in 32 tiles, connected by a 2D all-to-all mesh interconnect. The synchronization occurs in multiple stages, whereby a pair of threads notify each other of their arrival via atomic read and writes to synchronization variables. The architecture enforces cache coherence and the synchronization variables must be aligned to the cache boundary to avoid false sharing. The results indicate that at high core count, in such a fragmented and tiled many-core system, the butterfly barrier ($\sim2500$ cycles) outperforms less hierarchical centralized barriers ($\sim6500$ cycles) where a single master thread is responsible for verifying the \glspl{pe}' arrival. 
In \cite{Gao_2021_CLUSTER}, tree barriers are tested on different many-core architectures with up to \num{64} ARMv8 cores. Authors focus on tournament barriers with a tree structure that fit the hierarchical core-cache organization of the underlying architecture, achieving a synchronization overhead of \num{642.5}, \num{618.2}, and \num{356} kilo-cycles on the \textit{ThunderX2}, the \textit{Pythion 2000+}, and the \textit{Kunpeng920} processors, respectively. 
In \cite{Villa_2008_CASES}, four barriers (single-master linear barrier, tournament tree barrier, butterfly barrier, and all-to-all barrier) are tested on the model of a scalable architecture with $N_{PE}$ simple \glspl{pe} connected together with a highly scalable \gls{noc} interconnect. Barriers were tested on different network topologies for up to \num{128} \glspl{pe}. In highly connected \gls{noc} topologies, such as the \textit{Torus}, barriers have less than \num{400} cycles synchronization overhead, and the all-to-all barrier, where any \gls{pe} can simultaneously send a message to other \glspl{pe}, performs the best. In a \textit{Mesh} \gls{noc}, synchronization overhead takes up to \num{10000} cycles for the all-to-all barrier but is limited to less than \num{1000} cycles for the tree and the butterfly barriers, as tree barriers become optimal when interconnect resources are reduced.
In~\cite{Glaser_CoRR_2020}, the low core count (only \num{8} to \num{16} cores) of the multicore architecture makes hardware support for quadratic-complexity \gls{pe} to \gls{pe} signaling feasible and highly energy efficient. For the barriers implemented, the average length of periods where \glspl{pe} can work independently from each other with 10\% overhead of barriers corresponds to just 42 cycles. 

In the context of \gls{gpu} programming, the \gls{cuda} \gls{api} has primitives that enable cooperative parallelism, including producer-consumer parallelism and synchronization across a thread group, the entire thread grid or even multiple \glspl{gpu}~\cite{CUDABArriers_2017}. From the hardware viewpoint, NVIDIA \textit{Ampere} first added asynchronous barriers, with a non-blocking arrival phase that allows threads to keep working on independent data while waiting. When eventually all threads need data produced by others, they wait until everyone arrives, spinning on a shared variable. In \textit{Hopper} \glspl{gpu}, threads sleep instead of spinning while waiting, and transaction barriers are introduced to enable asynchronous copies of data~\cite{Hopperarchitecture_NVIDIA_2022}. Due to the proprietary nature of these tools, the behavior in hardware and the software implementation of \gls{gpu} barriers is not entirely transparent.

\begin{figure}[h!]
\centering
\includegraphics[width=\columnwidth]{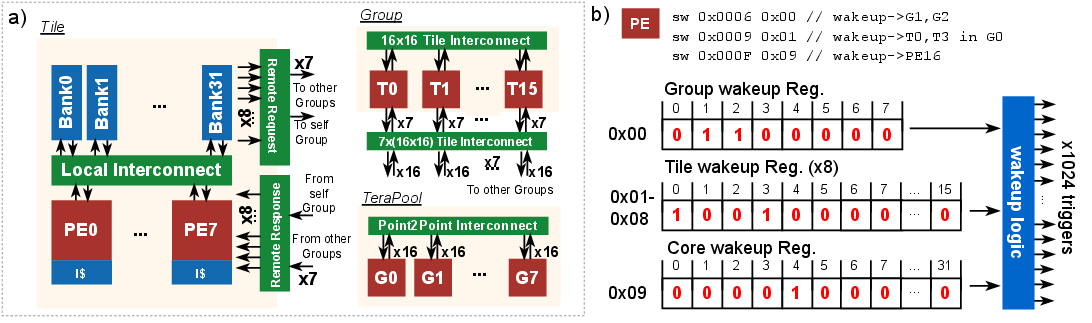}
\caption{a) The architecture of the \textit{TeraPool} cluster and b) scheme of the implemented wakeup cluster registers and triggers.}
\label{fig_architecture}
\end{figure}

\textit{TeraPool} is the scaled-up version of the \textit{MemPool} cluster presented in \cite{Mempool_DATE_2020, Mempool_Journal_2023}. The \textit{TeraPool} cluster, represented in Fig.~\ref{fig_architecture}, has \num{1024} \glspl{pe} tightly coupled to a multi-banked shared data memory. The access latency to any memory location is limited to \num{5} cycles, thanks to a hierarchical partition of the \glspl{pe} and of the AXI interconnection resources. \num{8} \glspl{pe} are grouped in a \textit{Tile}, with single-cycle access latency to \num{32} local banks via a local interconnect. \num{16} Tiles are part of a \textit{Group}. Each \gls{pe} in a Tile can access the memory banks of another Tile in the same Group in less than \num{3} cycles, thanks to the Group-level 16x16 interconnect. \num{8} Groups build a cluster, and each \gls{pe} can access a bank of a Tile in another Group in less than \num{5} cycles, though point-to-point connections between Groups. In a Tile, \glspl{pe} share the interconnection resources towards another Tile in the same Group, and in a Group, Tiles share the interconnection resources towards another Group. Contentions may arise when more than one memory request tries to access the same memory bank or the same shared interconnection resource, leading to one of them being stalled for one cycle. The cluster has a \gls{numa} interconnect, but the low access latency in the case of no contentions makes \textit{TeraPool} a good approximation of the \gls{pram} model~\cite{PRAM_JaJa_2011}. 

In \cite{Mohamed_2023_Wiley, Gao_2021_CLUSTER, Villa_2008_CASES} the best performance of a barrier algorithm over another is strongly dictated by the interconnection topologies and by the clustering of \glspl{pe} in hierarchies. In \textit{TeraPool}, synchronization variables in any bank of the shared memory can be accessed at an almost equivalent and low cost, allowing to choose the granularity of barrier  synchronization on the basis of the workload characteristics rather than of the topology of the interconnection between \glspl{pe} and of the hierarchical partition of the hardware.

\begin{figure}[h!]
\centerline{\includegraphics[width=0.9\columnwidth]{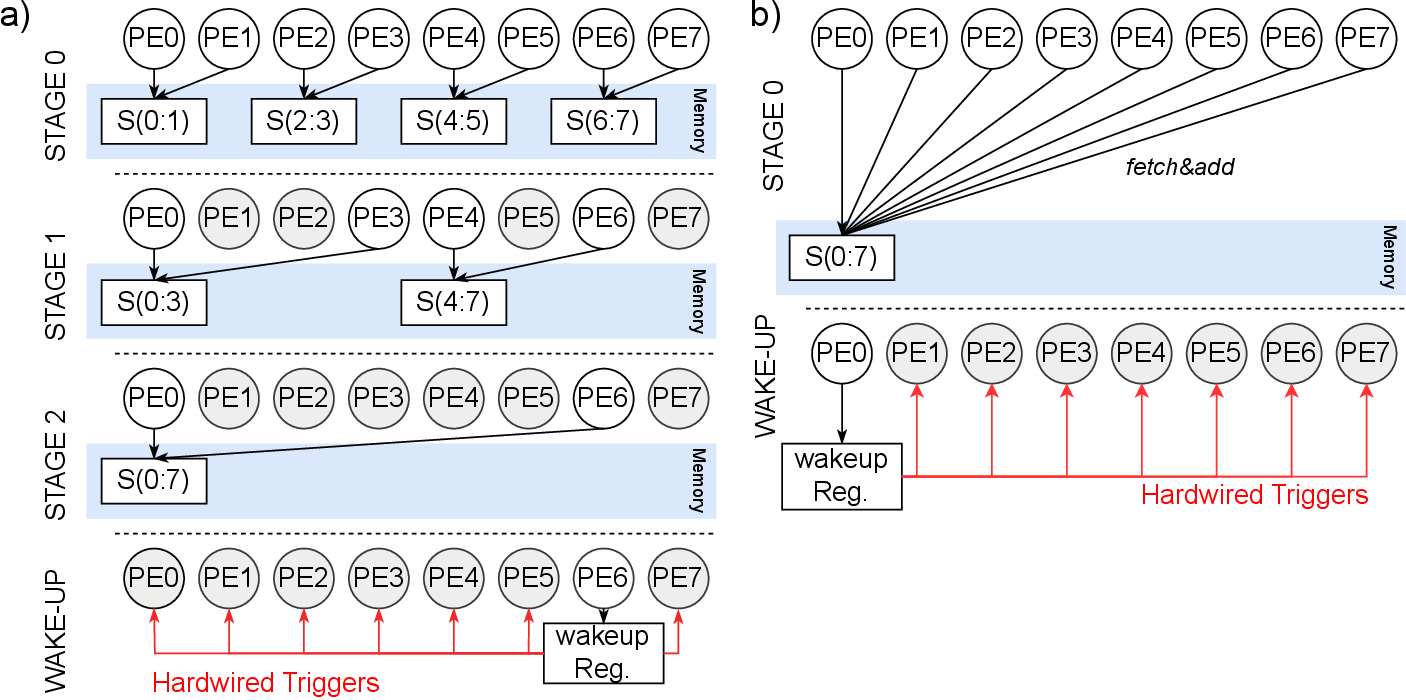}}
\caption{a) Binary tree for the arrival phase of the barrier. Couples of \glspl{pe} synchronize by atomically accessing shared synchronization variables. b) Central counter barrier.}
\label{fig_karytree}
\end{figure}

\section{Barriers implementation}
In this section, we describe the implementation of our barrier primitives. A synchronization barrier algorithm can be divided into three phases: an arrival, a notification, and a re-initialization phase.

For the arrival phase, we adopt a k-ary tree. The $N_{PE}$ cores of the cluster are divided into $N_{PE}/k$ groups of $k$ \glspl{pe}. In each group, synchronization occurs in the form of a central-counter barrier~\cite{Hoefler_2004}. Each \gls{pe} arriving at the barrier updates a shared counter via an atomic \textit{fetch\&add} operation and goes into a \gls{wfi} sleeping state. The last \gls{pe} reaching the synchronization step, fetches a counter value equal to $k-1$ and continues with the next steps, where it is further synchronized with the other $N_{PE}/k - 1$ \glspl{pe} that survived the first step. The last step counts $k$ \glspl{pe}, and the very last \gls{pe} arriving wakes up all the \glspl{pe} in the cluster. The arrival tree works best when the $log_k(N_{PE})$ is an integer, but it is also adapted to the case where $k$ is any power of \num{2} $<N_{PE}$, by synchronizing a number of \glspl{pe} different from the radix of the tree in the first step of the barrier. Varying $k$, we encounter two extremes, represented in Fig.~\ref{fig_karytree} (a-b): the left shows a radix-2 logarithmic tree barrier, where each step only synchronizes pairs of \glspl{pe}, the right illustrates the central-counter barrier. The re-initialization phase is implemented concurrently with the arrival phase, as each \gls{pe} arriving last in a synchronization step also resets the shared barrier counter before proceeding to the next step.

The notification phase leverages hardware support in the form of a centralized wakeup handling unit. The last \gls{pe} arriving at the barrier and fetching from the shared barrier counter variable writes in a cluster shared register. The address of this register is in the cluster global address space and can be accessed by any core through the hierarchical AXI interconnections. The written value is detected by the wakeup handling logic that sends a wakeup signal to each individual \gls{pe}, triggering $N_{PE}$ hardwired wakeup lines. A software implementation of the wakeup mechanism is excluded because it would fall into the single master barrier class~\cite{Mohamed_2023_Wiley}, whose cost scales linearly with $N_{PE}$ and is unsuitable for synchronizing more than a few tens of \glspl{pe}.

We support synchronizing a subset of \glspl{pe} in the cluster modifying the wakeup handling unit by adding other memory-addressable shared registers, as shown in Fig.~\ref{fig_architecture}. The core wakeup register is a 32-bit register that can be used to either trigger a wakeup signal to all the \glspl{pe} in the cluster, when it is set to all ones, or to a single \gls{pe}, by specifying its ID. One 8-bit register is used to selectively wake up Groups, and a register per Group is added to wake up Tiles in a Group selectively. A bitmask is used to determine the Groups or the Tiles to wake up. Depending on the bitmask written by a \gls{pe} in one of the synchronization registers, the wakeup logic, asserts a subset of or all the wakeup triggers hardwired to the cores in the cluster, to trigger a wakeup signal.

The implemented barriers can be called from the function body through a custom software \gls{api}. The radix of the barrier can be tuned through a single parameter, to ease trials and selection of the best synchronization option.

\section{Benchmarking strategy}
In the following, we describe the benchmarking strategy adopted to evaluate the performance of our barriers. Software is compiled with GCC 7.1.1 and runs on the open-source \gls{rtl} model of \textit{TeraPool}, via a QuestaSim 2021.2 cycle-accurate simulation. In all the cases we assume that the input data resides in the L1 memory of the cluster.

\subsection{Benchmarking with random delay}
We first test the implemented barriers on the synchronization of \glspl{pe} with a synthetic kernel implementing a random execution time for the parallel threads. The \glspl{pe} start the execution together and proceed in parallel through a synchronization-free time interval. At this point, before entering the barrier, the cores are delayed by a number of cycles drawn from a uniform distribution between zero and a maximum delay. We track the average time spent by the \glspl{pe} in the barrier. Since the result is subjected to the randomness of the delay, we average it over multiple tests. We also compute the fraction of the cycles spent in a barrier over the total runtime, as a function of the initial parallel section of the program only, referred to as \gls{sfr}. The goal is to estimate the minimum \gls{sfr} for a negligible overhead, which is important information for the programmer on the granularity of the workload allocation to \glspl{pe}.

\subsection{Benchmarking of kernels}
We analyze the performance of the barriers on benchmark kernels with a key role in the field of signal processing and telecommunications. The kernels are implemented using a fork-join programming model, in which each \gls{pe} operates on separate portions of the input data and accesses memory concurrently with the others. The final synchronization is achieved through a barrier call after all \glspl{pe} completed their tasks. We can identify three classes of kernels: 

\begin{itemize} {
\item The \gls{dotp} and the \gls{axpy} are implemented enforcing local access of the \glspl{pe} to memory so that all the inputs can be fetched with one cycle latency. The data to be processed is equally divided between the \glspl{pe}. The dot-product implies a reduction, which is implemented via the atomic addition of each \gls{pe}'s partial sum to a shared variable.
\item In the \gls{dct} and \gls{gemm} the workload is equally distributed between \glspl{pe}, but we cannot enforce local access for all the \glspl{pe}. Therefore, we expect some memory requests to take on average more cycles because of the inherent interconnection topology and contentions arising from concurrent parallel accesses.
\item In the \gls{conv2D}, the workload is not equally distributed because some \glspl{pe} are just used to compute the image border, but the access pattern is locally constrained as seen for \gls{axpy} and \gls{dotp}.
} \end{itemize}

\begin{figure}[h!]
\centerline{\includegraphics[width=0.9\columnwidth]{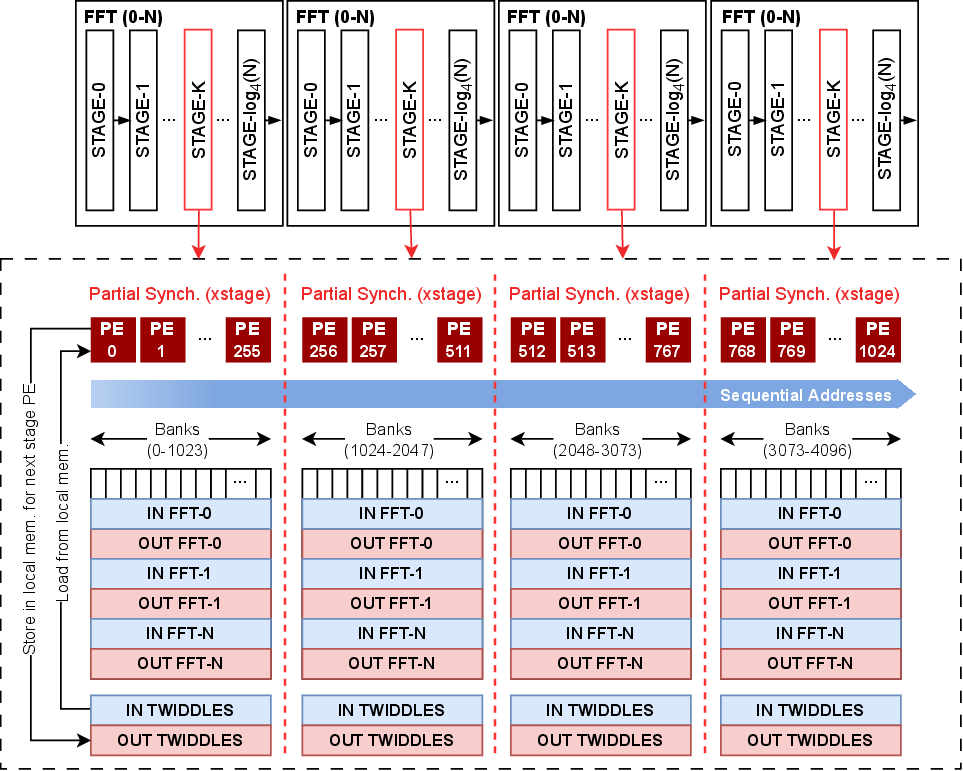}}
\caption{Scheduling of $N\times4$ FFTs on \textit{TeraPool}. Stage by stage, a group of \textit{N} independent FFTs is scheduled on the same subset of \gls{pe}, which are then partially synchronized.}
\label{fig_fftscheduling}
\end{figure}

\subsection{Benchmarking of a 5G-processing application}
Finally, we consider the implementation and performance of a full 5G application running on \textit{TeraPool}, evaluating the impact of synchronization overhead on parallelization. We benchmark the \gls{ofdm} demodulation stage of the lower \gls{phy} in 5G \gls{push}, followed by a digital beamforming stage. In this workload, $N_{RX}$ independent antenna streams, consisting of $N_{SC}$ orthogonal sub-carrier samples, undergo a \gls{fft}. The beamforming process is a linear combination of the antennas' streams, by known coefficients, producing $N_{B}$ streams of $N_{SC}$ samples. Between the two steps, there is a strong data dependency, and synchronization is needed.

The implemented radix-4 decimation in frequency \gls{fft} is a challenging multi-stage synchronization kernel, because \glspl{pe} need to be synchronized after each butterfly stage. In a butterfly stage, each \gls{pe} combines \num{4} inputs to produce \num{4} outputs. We can store the inputs locally in the same bank on different rows. Given the banking factor of \num{4}, each core can work with the minimum access latency on $4\times4$ input data and store the data in the local banks of \glspl{pe} that will use them in the next \gls{fft} stage. A 4096-points \gls{fft} is therefore stored in the local memory of \num{256} \glspl{pe} and processed by them, as shown in Fig.~\ref{fig_fftscheduling}. Synchronization overhead is kept to the bare minimum: every \gls{fft} stage is run in parallel over a subset of \glspl{pe} of the cluster. Cores working on different FFTs are independently synchronized, leveraging the partial barriers. Since a barrier is needed after each \gls{fft} computation stage, the \glspl{pe} can load twiddles and work on multiple independent \glspl{fft} before joining, thus reducing the fraction of synchronization overhead over the total runtime.

Beamforming is implemented as a \gls{gemm} between the $N_B\times N_{RX}
$ matrix of beamforming coefficients and the output \gls{fft} streams. Each \gls{pe} computes a different output element, as the dot-product between a row and a column of the first and second input matrix respectively. The workload is distributed column-wise between \num{1024} \glspl{pe}, so that each column goes to a different \gls{pe}, while accesses to rows can happen concurrently. 

We try different configurations, as proposed in the \gls{3gpp} technical specifications~\cite{3gpp.38.214}. Assuming $N_{SC}=4096$, we consider a \gls{mimo} systems with $N_{RX}=16$, \num{32}
or \num{64} antennas and $N_B=32$ beams.

\section{Results}
\begin{figure}[h!]
\centering
\includegraphics[width=\textwidth]{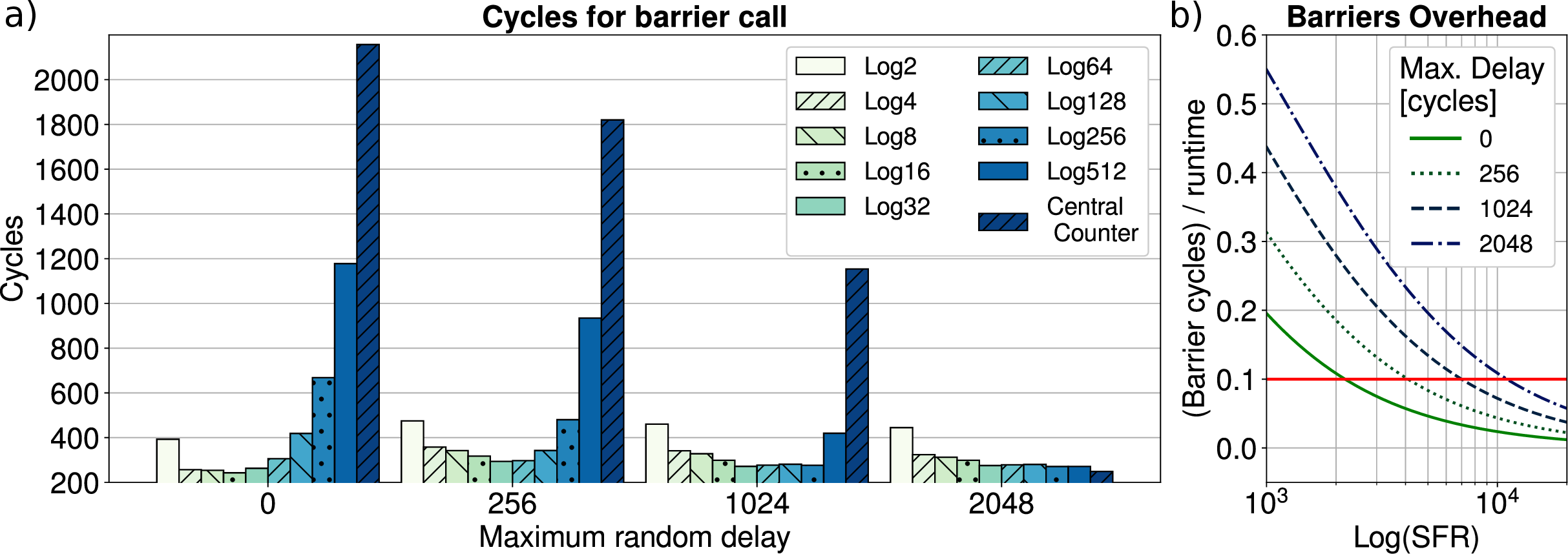}
\caption{a) Cycles between the last \gls{pe} entering and leaving the barrier for different barrier radices, and maximum delays between the incoming \glspl{pe}. b) Fraction of the best-performing barriers' overhead as a function of the \gls{sfr}.}
\label{fig_cyclesbarrier}
\end{figure}

This section discusses the results of our benchmarking experiments. Fig.~\ref{fig_cyclesbarrier} (a) represents the cycles between the last \gls{pe} entering and the last one leaving the barrier when each \gls{pe} has a different random delay, extracted from a uniform distribution between zero and a maximum value. When the delay is zero for all \glspl{pe}, the cycles for the barrier call exhibit a scoop behavior depending on the radix used. Low radix barriers require a longer multi-step synchronization process. In this sense, the binary-tree logarithmic barrier is the worst, having \num{10} steps where \glspl{pe} are synchronized in pairs. Barriers with few centralized synchronization variables require simultaneous access to the same memory locations by multiple \glspl{pe}, creating banking conflicts. On this side, the linear central-counter barrier, where \num{1024} \glspl{pe} conflict for the same memory location, is the worst. 
As the maximum delay increases, the \glspl{pe}' arrival at the barrier is scattered. Therefore, contentions in accessing the synchronization variables reduce, and the lower radix barriers start to be more expensive than the higher radix ones, generating a staircase pattern. Ultimately, for the \num{2048} cycles delay, the central-counter barrier is the best, because cores requests to the barrier variable are sufficiently scattered in time to avoid contentions. The time that the last \gls{pe} arriving at the barrier needs to traverse all the levels of the k-ary tree is, in this case, the most relevant part of the barrier runtime.

\begin{figure}[h!]
\centering
\includegraphics[width=\textwidth]{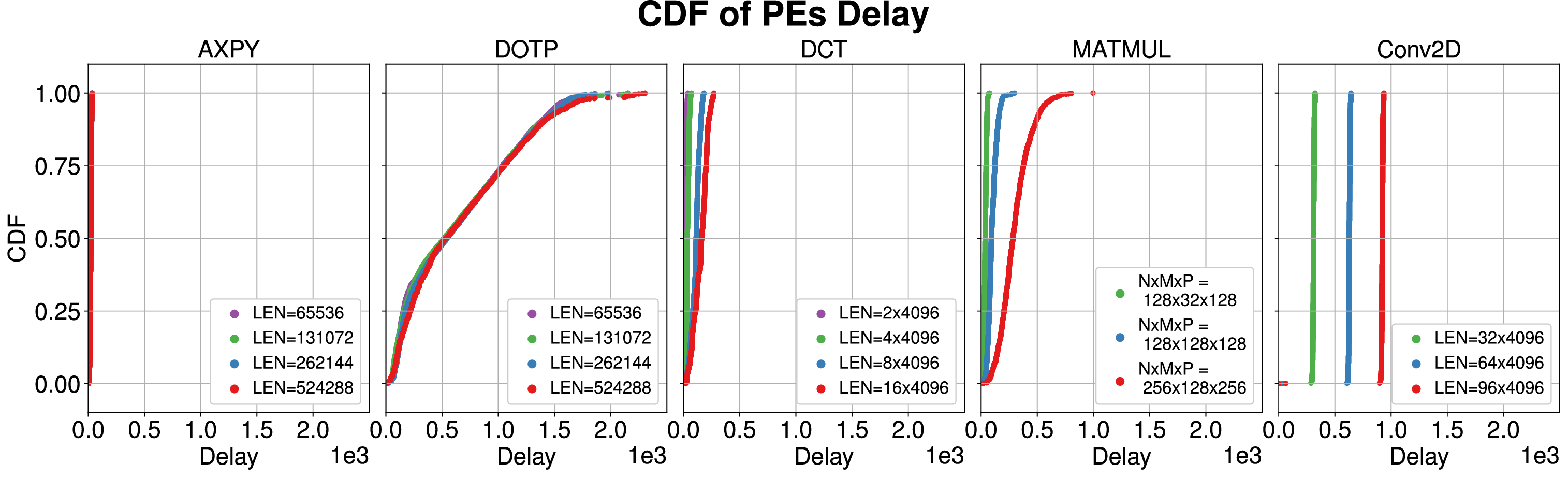}
\caption{CDF of the distributions of the difference between the fastest and the slowest \gls{pe} runtimes for different exemplary parallel kernels before synchronization.}
\label{fig_cdfkernels}
\end{figure}

Fig.~\ref{fig_cyclesbarrier} (b) shows the fraction of the average cycles spent in a barrier by a \gls{pe}, over the total runtime cycles, as a function of the \gls{sfr}. The \glspl{pe}' arrival is scattered as in the previous experiment. We consider different values of the maximum random delay between the \glspl{pe}, and for each case, we report the results on the barrier with the best-performing radix. To achieve a synchronization overhead of less than 10\%, our barriers need a \gls{sfr} between \num{2000} and \num{10000} cycles, depending on the scattering of the arriving \glspl{pe}.

Since the arrival time of \glspl{pe} is not always uniformly distributed, we measured the actual distribution for various key kernels and evaluated its impact on synchronization.
Fig.~\ref{fig_cdfkernels} represents the \gls{cdf} of the difference between the runtime cycles of the fastest and the slowest \gls{pe} before synchronization for different parallel kernels.
\begin{itemize} {
\item The local access enforced for the \gls{axpy} and the \gls{dotp} kernels makes them conclude their job at the same time, independently of the input dimension. Contentions in accessing the reduction variable make some \glspl{pe} slower than others for the execution of the \gls{dotp}.
\item The access latency and the contentions in fetching memory locations that are distributed over the banks make the arrival of the \glspl{pe} executing \gls{dct}, \gls{gemm}, and \gls{conv2D} more scattered. The difference between the arrival time of \glspl{pe} depends on the input dimension, because the larger the input data, the more the contentions. Interestingly the most compact distribution in arrival times for the \gls{dct} kernel is obtained for an input of length $2\times4096$. In this case, each \gls{pe} works on two inputs of $2\times2$ samples. Since \textit{TeraPool} has \num{1024} \glspl{pe} and a banking factor of \num{4}, and the addresses run sequentially, the data is always stored locally.
\item In the case of the \gls{conv2D}, we see a wide gap between the first and the last \glspl{pe} arriving, which is caused by work imbalance. Some \glspl{pe} are indeed assigned to the calculation of the boundary of the input image, which containing zeros is resolved in a lower number of instructions with respect to the pixels in the center of the image. 
}\end{itemize}

\begin{figure}[h!]
\centering
\includegraphics[width=\textwidth]{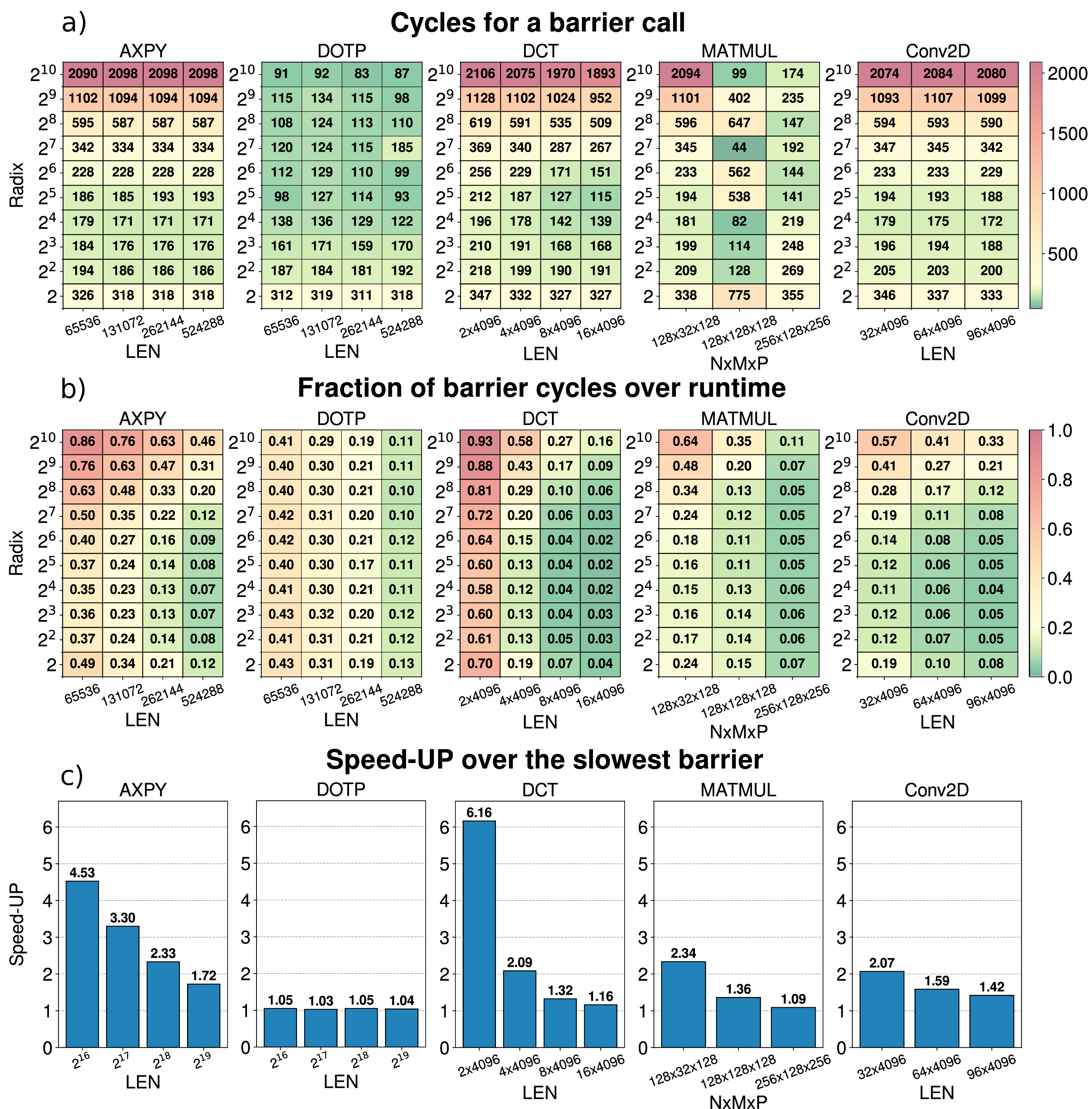}
\caption{Colormaps of a) the delay between the last \gls{pe} leaving and the last \gls{pe} entering the barrier, b) the fraction of cycles for a barrier call (average on all \glspl{pe}) over the total runtime, for different kernels, input dimensions, and barrier radices. c) Speed-up of the kernel synchronized with the fastest barrier on the kernel using the slowest barrier for each input dimension.}
\label{fig_cmapskernels}
\end{figure}

\begin{figure}[h!]
\centering
\includegraphics[width=\columnwidth]{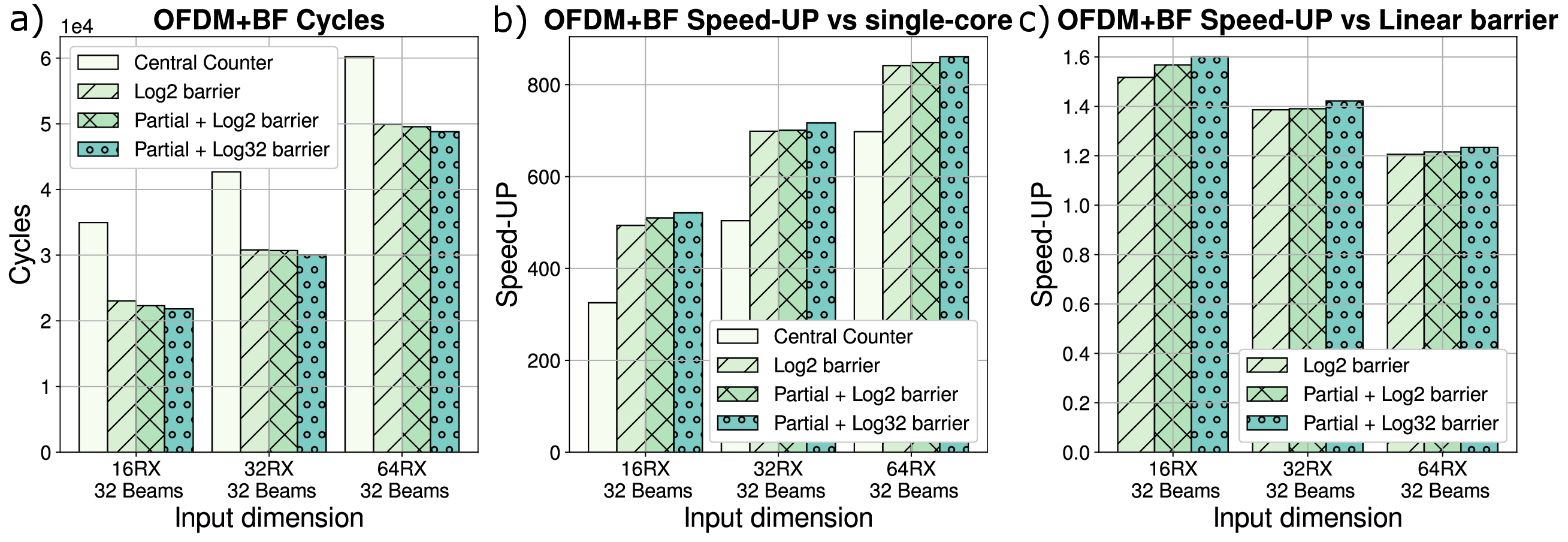}
\caption{a) Execution cycles for the 5G \gls{ofdm} and beamforming with different synchronization barriers. b) Speed-up with respect to serial execution on a Snitch core. c) Speed-up with respect to the baseline using the linear central-counter barrier.}
\label{fig_ofdmresults}
\end{figure}

Fig.~\ref{fig_cmapskernels} (a) reports the delay between the last \gls{pe} leaving the barrier and the last \gls{pe} entering the barrier for different exemplary kernels, input dimensions, and radices of the k-ary tree barrier. The \gls{axpy} and the \gls{dct} find their sweet spot around radices \num{32} and \num{16} because the scattering in the arrival of \glspl{pe} is moderate and the use-case falls in the scoop region on the plot in Fig.~\ref{fig_cyclesbarrier} (a). The arrival of the \glspl{pe} in the \gls{conv2D} kernel is split between the fast cores computing the border of the input image and the slow cores computing the inner pixels, which are predominant, resulting in similar behavior to \gls{axpy} and \gls{dct}. The arrival of the \glspl{pe} working on the \gls{dotp} kernel is scattered because of the reduction process. In this case, the lower-radix barriers have the worst performance, and the central counter barrier shows the best performance. We identify a behavior that is close to the staircase pattern on the right-hand side of Fig.~\ref{fig_cyclesbarrier} (a). The sparsity in the arrival of \glspl{pe} strongly depends on the input dimension for the \gls{gemm} kernel. Therefore, we find a small delay behavior for the very steep distribution obtained in the case of the input $128\times32\times128$ and a large delay behavior for the smooth distribution obtained in the case of the input $256\times128\times256$. The intermediate dimension has an intermediate behavior with some outlier points, caused by a peculiar feature of our barrier implementation: the synchronization of subsets of \glspl{pe} in the leaves nodes of the tree barriers is initially performed on \glspl{pe} with a contiguous index, therefore, mostly between \glspl{pe} in the same Tile or in the same Group. These \glspl{pe} share the interconnection resources, which are under stress during the execution of the \gls{gemm}, and have a delay between them that slows down the first phase of the tree synchronization.

Fig.~\ref{fig_cmapskernels} (b) shows the average over all the \glspl{pe} of the ratio between the barrier cycles and the total runtime for the selected kernels. The \gls{axpy} has a hot loop with few operations per input. Therefore, the problem dimension to achieve $\sim10\%$ overhead of the barriers is large. The \gls{dotp}, the \gls{gemm}, and the \gls{conv2D} having a higher ratio of computations per input and steep \glspl{cdf} of the \glspl{pe}' arrival times, benefit of our tree barriers from small data dimensions. The \gls{sfr} of the \gls{dotp} is large due to the \gls{pe} scattering produced by reduction. This also requires large input vectors to make the barrier overhead negligible. 

Fig.~\ref{fig_cmapskernels} (c) reports the speed-up on the total runtime obtained synchronizing with the fastest barrier option for every input dimension and kernel, compared to the slowest. Speed-up decreases as the input dimension grows because the barrier fraction on the total runtime reduces. For the \gls{dotp} speed-up is limited, because, as clearly shown on the right-hand side of Fig.~\ref{fig_cyclesbarrier}, the gap between barriers is small when cores arrive scattered at synchronization. For all the other kernels, even in the case of large inputs, when synchronization consists of less than 10\% of the total runtime, we report speed-up between $1.1\times$ and $1.72\times$. This analysis proves that choosing the barrier radix based on the parallel kernel characteristics provides significant advantages.

Finally, in Fig.~\ref{fig_ofdmresults} we compare the performance of the central counter barrier, the tree barriers, and the partial barriers on the 5G \gls{ofdm} and beamforming workload. Overheads account for the multi-stage synchronization required by \gls{fft} and the synchronization enforced by data dependencies between \gls{fft} and \gls{gemm}. Fig.~\ref{fig_ofdmresults} (a) shows the execution cycles of the 5G application under exam for different numbers of antenna streams, and Fig.~\ref{fig_ofdmresults} (b) reports the speed-ups with respect to a sequential execution on a single Snitch core. The radix-4 \num{4096} points \glspl{fft} are scheduled on the cluster as described in Fig.~\ref{fig_fftscheduling}. We notice that using tree barriers greatly reduces the runtime with respect to using a central counter barrier, improving the serial speed-up. Constraining memory access to local addresses avoids a scattered arrival of the cores due to contentions, which explains the benefit obtained from tree synchronization. 
Further improvement can be obtained by fine-tuning the radix of the tree barrier through the provided \gls{api} and by introducing partial synchronization. The speed-ups with respect to synchronization using a central-counter barrier are reported in Fig.~\ref{fig_ofdmresults} (b). The best result of $1.6\times$ is obtained using a radix-32 barrier and synchronizing only groups of \num{256} \glspl{pe} working on independent \glspl{fft}. The overall speed-up reduces as the number of independent \glspl{fft} run between barriers increases. The inefficiency of the central-counter barrier slowly fades away as synchronization becomes a negligible part of the runtime. On our best benchmark, corresponding to the run of $4\times16$ \glspl{fft} of \num{4096} points and of a \gls{gemm} between a $32\times64$ and a $64\times4096$ matrix, the synchronization overhead accounts for just 6.2\% of the total runtime. In this last case, we observe a speed-up of $1.2\times$.

\section{Conclusions}
This work challenged the fork-join programming model by scaling it to the \num{1024} cores of the \textit{TeraPool} architecture. To leverage \textit{TeraPool}'s full parallel workforce for parallel workloads, we focused on optimizing a key synchronization primitive: the shared memory barrier. We developed and optimized different software implementations of these barriers and added hardware support for the partial synchronization of groups of \glspl{pe} in the cluster. We then tested our synchronization primitives on kernels of paramount importance in the field of signal processing. The average of the cycles spent by cores in a barrier over the total runtime cycles for the \gls{axpy}, \gls{dotp}, \gls{dct}, \gls{gemm}, and \gls{conv2D} kernels is respectively as low as \num{7}\%, \num{10}\%, \num{2}\%, \num{5}\%, and \num{4}\%. On the sequence of \gls{ofdm} and beamforming, a full 5G processing workload, which includes a multi-stage synchronization kernel (\gls{fft}), logarithmic tree barriers and the use of partial synchronization outperform the central-counter synchronization with a $1.6\times$ speed-up. Our scheduling policy allows reducing the fraction of synchronization cycles over the total cycles to less than 6.2\%.

Our results demonstrate that despite its high core count, the \textit{TeraPool} many-core cluster behaves as a good approximation of the \gls{pram} model with low synchronization overhead. This result relies on a tuned selection of the barrier flavor for a given target kernel.

\subsubsection{Acknowledgements} This work was supported by Huawei Technologies Sweden AG.

\bibliographystyle{splncs04}
\bibliography{refs.bib}
\end{document}